# Low-temperature synthesis of high quality Ni-Fe layered double hydroxides hexagonal platelets

Sonia Jaśkaniec,[a,b] Christopher Hobbs,[b,c] Andrés Seral-Ascaso,[b,c] João Coelho,[b,c] Michelle P. Browne,[d] Daire Tyndall,[a,b] Takayoshi Sasaki[e] and Valeria Nicolosi[a,b,c]

[a] School of Chemistry, Trinity College Dublin, Ireland, [b] CRANN&AMBER, Trinity College Dublin, Ireland, [c] School of Physics, Trinity College Dublin, Ireland, [d] School of Chemistry, Queens University Belfast, Ireland, [e] National Institute for Materials Science, Tsukuba, Japan

This paper describes the wet-chemistry synthesis of highly crystalline hexagonal flakes of Ni-Fe layered double hydroxide (LDH) produced at temperature as low as 100 °C. The flakes with diameter in the range of 0.5-1.5 μm and the thickness between 15 and 20 nm were obtained by homogeneous precipitation method with the use of triethanolamine (TEA) and urea. By analyzing the intermediate products, it is suggested that, differently from previous reports, a thermodynamically metastable iron oxyhydroxide and Ni-TEA complex are firstly formed at room temperature. Subsequently, when the mixture is heated to 100 °C and the pH increases due to the thermal decomposition of urea, $Ni^{2+}$ and $Fe^{3+}$ are slowly released and then recombine, thus leading to formation of pure, highly-crystalline Ni-Fe LDH flakes. This material showed promising results as an electrocatalyst in oxygen evolution reaction (OER) providing an overpotential value of 0.36 V.

## Introduction

Ni-Fe LDH layered double hydroxide (LDH) have become the focus of an extensive scientific research due to the high electrocatalytic activity of this material in oxygen evolution reaction (OER).[1-3] LDH have a structure similar to brucite where single layers are made of edge sharing $Mg(OH)_6$ octahedrons.[4] Contrary to brucite, the layers of LDH present a net positive charge due to the partial substitution of divalent cations with trivalent ones.[5] In order to ensure overall electrical charge neutrality, anions are intercalated within the interlayer space, which results in electrostatic interactions between the layers. These forces together with van der Waals interactions and hydrogen bonds keep the sheets together to form a three-dimensional layered framework.[5,6]

Recently it was suggested that Ni-Fe LDH catalytic activity is related not only to the surface area, but also to the number of open coordination sites at the edges,[7] so the design of simple synthesis routes for the preparation of crystalline, thin platelets with homogeneously distributed Ni and Fe is crucial in order to achieve reliable and highly active catalysts.[3] However, the synthesis of $Fe^{3+}$-containing LDH with high crystallinity and a well-defined shape is challenging because usually gel-like, water-insoluble $Fe(OH)_3$ precipitates at pH above 2, which impedes the further incorporation of $Ni^{2+}$ within its structure.[8-10] From several techniques which have been used to obtain Ni-Fe LDH in recent years (such as reversed co-precipitation,[11] ball milling,[12] topochemical routes[8,13] or co-precipitation with the use of long-chain organic acid[14]), special attention has been placed on the use of capping agents, such as trisodium citrate or triethanolamine (TEA), which coordinate $Fe^{3+}$ thus preventing the formation of its hydroxide while at the same time facilitating the combination with $Ni^{2+}$.[9,10]

The synthesis of Ni-Fe LDH with the use of TEA follows the so-called "atrane route".[9] In a simple way, the formation of Fe-TEA complexes, which are inert to hydrolysis, prevents the precipitation of insoluble iron hydroxides. This inertness is controlled by pH and temperature, so when both parameters are increased,[4] Fe-TEA complex decomposes and $Fe^{3+}$ ions recombine with other metal ions present in the solution to form mixed-metal LDH.[9] However, the hydrolysis of this atrane complex takes place at temperature exceeding the boiling point of the solvent (typically 150 °C),[3,9] which, in case of wet-chemistry synthesis, involves the use of autoclaves. In general, the scalability of the hydrothermal methods is limited by the size of a pressure vessel and the total synthesis costs are significantly higher, which results from special equipment and high temperature requirements.[9]

In this work, pure, highly crystalline Ni-Fe LDH hexagonal platelets were produced by homogeneous precipitation at a relatively low temperature of 100 °C. In contrast to previous reports,[3,9] we used a large excess of the capping agent in relation to $Fe^{3+}$ ions which caused the precipitation of a metastable solid iron(III) oxyhydroxide at room temperature, while nickel(II) remained in the solution coordinated by TEA molecules. Then, upon heating and pH increase the reaction intermediates decompose and recombine forming Ni-Fe LDH. This process was carried out in a conventional round-bottom flask, suggesting the possibility of being easily scaled up. The material was tested as electrocatalyst in OER showing activity typical of Ni-Fe LDH nanoparticles.

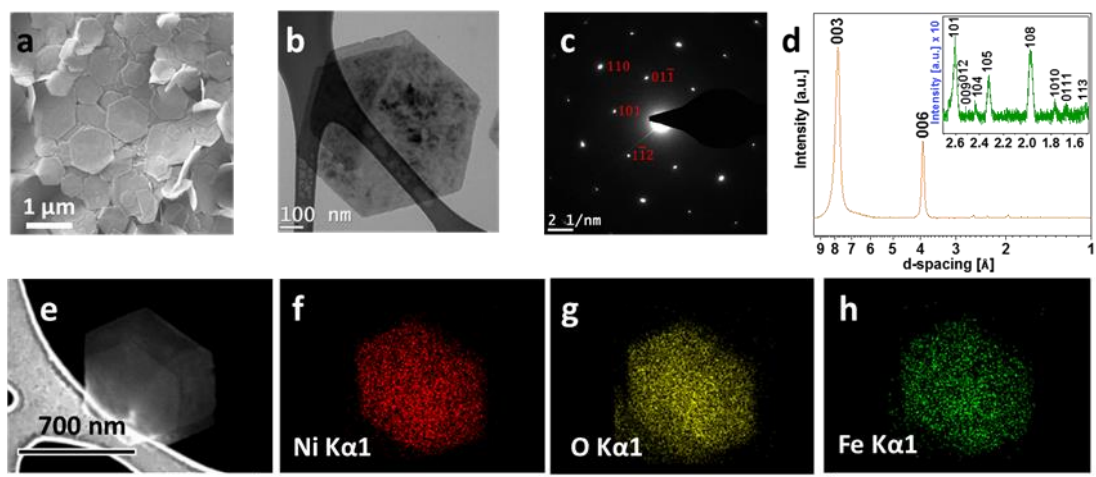

Fig. 1 SEM and TEM micrograph (a, b); SAED pattern (c); XRD pattern (d); SEM- EDX elements mapping (e-h) of the Ni-Fe LDH hexagonal platelets.

## Results and discussion

*The morphological and structural characterization of Ni-Fe LDH hexagonal platelets*

The morphological and structural characterization of the as-prepared LDH was carried by SEM, TEM, AFM, FT-IR and TGA analysis. SEM micrographs (Fig. 1a) revealed that the as synthesized sample is composed of flakes with a well-defined hexagonal shape and on the order of 1 μm in lateral dimension. The height profiles of the flakes, measured by AFM (Fig. 2), are in the range of 15 to 20 nm, which roughly corresponds to 18-25 layers.

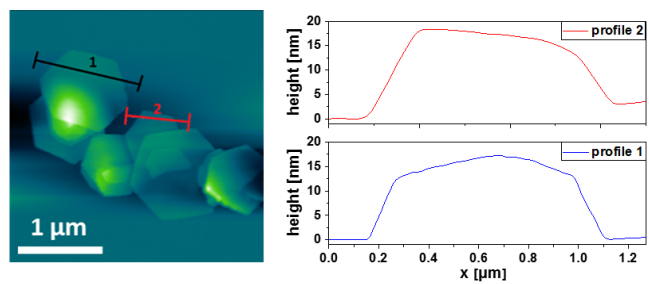

Fig. 1 AFM image and height profiles of Ni-Fe LDH hexagonal platelets.

The homogenous contrast observed in the TEM micrographs (Fig. 1b) suggests that the flakes have a uniform thickness and the SAED pattern (Fig. 1c) shows a hexagonal symmetry (a=3.08 Å, c= 23.55 Å), which is confirmed by XRD patterns (Fig. 1d). The absence of additional reflections in the SAED pattern indicates that the sample is solely composed by Ni-Fe LDH. This is evidenced by the (101), ($01\bar{1}$), ($1\bar{1}2$) and (110) lattice planes (Fig. 1c).

The homogeneous distribution of the metal cations within the nanoflakes[15] was confirmed via SEM-EDX mapping (Fig. 1e-h). Nickel, iron and oxygen are evenly distributed in the flakes surface, thus indicating a consistent substitution of $Ni^{2+}$ by $Fe^{3+}$ in the brucite-like layers of $Ni(OH)_2$.

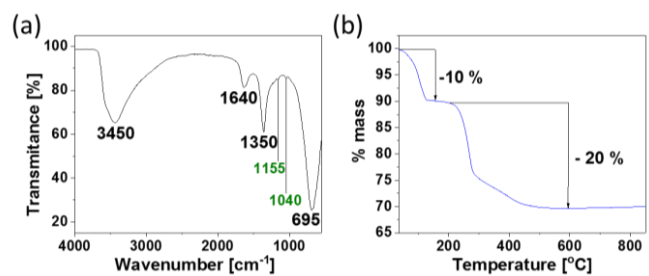

Fig. 2 FT-IR spectrum (a) and TGA curve (b) of Ni-Fe LDH flakes.

FT-IR analysis of the flakes (Fig. 3a) revealed the typical spectrum of LDH. Adsorbed water in the samples leads to the broad signal at 3450 $cm^{-1}$, which is characteristic of the O-H stretching mode in brucite-like layers. The O-H bending vibrations in water molecules is also present at 1645 $cm^{-1}$.[15, 16] A strong band at 690 $cm^{-1}$ is usually ascribed to the $\upsilon_{M-O}$ lattice vibrations.[17] The absorption band observed at 1350 $cm^{-1}$ is related to the stretching mode of $CO_3^{2-}$,[17] while the weak peaks at 1155 $cm^{-1}$ and 1040 $cm^{-1}$ (green labels) are characteristic of the C-N stretching mode in tertiary amines,[18] suggesting a small contamination with TEA.

LDH are known to thermally transform into mix-metal oxide and spinel phases.[19] The TGA curve (Fig. 3b) indicates that this process occurs in two steps.[9] Firstly, in the temperature range of 30 °C to 200 °C, the adsorbed and interlamellar water is removed, resulting in a 10 % mass loss. Secondly, at temperatures of 200 to 800 °C, the weight loss is promoted by the dehydroxylation of the brucite-like layers and the decomposition of the counter anions.[17] The total mass loss of roughly 30 % is consistent with other works reported on the thermal decomposition of LDH materials. Finally, $Ni^{2+}/Fe^{3+}$ ratio was calculated as 3.54 by atomic absorption spectroscopy which, in combination with TGA results, led to the formula $Ni_{0.78}Fe_{0.22}(OH)_2(CO_3)_{0.11}\cdot 0.6H_2O$.

## Electrocatalytic activity in OER

The performance of the $Ni_{0.78}Fe_{0.22}(OH)_2(CO_3)_{0.11}\cdot 0.6H_2O$ as an electrocatalyst for OER was tested by linear sweep voltammetry polarization curve (Fig. 4).

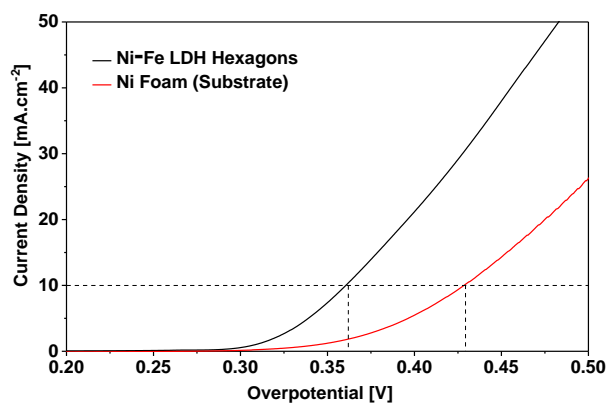

Fig. 3 Linear sweep voltammetry curves for the Ni-Fe LDH hexagons and Ni foam at a scan rate of 5mV·s$^{-1}$ in 1M KOH.

As expected the synthesized Ni-Fe LDH exhibits electrocatalytic activity towards OER, presenting an overpotential of 0.36 V, which is similar to other nanostructured Ni-Fe LDH materials.[2, 20-24] The overpotential of the support (Ni-foam) was also measured resulting in a value of 0.43 eV.

Variations in the OER overpotential among different studies is quite common and it is likely related to the difference in LDH thickness, the used substrate and the electrolyte solution concentration. In spite of presenting the same chemical composition of their bulk counterparts, monolayer LDH are always preferred as their catalytic activity is always optimized.[1] Higher electrolyte concentrations can also positively impact the OER activity of Ni-Fe LDH.[25] Currently, the state of the art catalysts rely on ultrathin two dimensional materials combined with carbon materials. The hybridization or direct mixing of LDH with carbon nanotubes, graphene, graphene oxide and quantum dots, compensates for the low electrical conductivity of the original materials. Therefore, the optimization of the OER catalytic properties of the LDH presented in this study would be considered for future work.

## Possible formation mechanism of Ni-Fe LDH platelets

Results presented so far indicate that homogeneous precipitation at low temperature is a very interesting technique for the synthesis of crystalline LDH hexagons with promising electrocatalytic properties. However, the processes behind this approach are not fully understood. Further experimental work allowed us to come forth with a hypothetical model for the LDH formation by homogeneous precipitation at low temperature. First, the addition of TEA to the reaction mixture drastically changes the pH of the starting solution (from 2.5 to 6.6). At these conditions, a gel-like brown precipitate was formed at room temperature (Fig. 5a), which upon heating to 100 °C and pH increase transformed to Ni-Fe LDH hexagonal platelets. In order to try to clarify how the final product is produced we split the synthesis process between two steps: at room temperature and at 100 °C.

### Step 1- Room Temperature

The formation of the iron oxyhydroxide precipitate at room temperature was investigated by UV-Vis measurements performed at 350 nm, which is linearly related to the Fe$^{3+}$ concentration (Ni$^{2+}$ has a negligible absorbance at that wavelength, Fig. S1 in ESI†). Before recording the UV-Vis spectra the reaction mixture was centrifuged (3000 rpm/10 min.) in order to remove the precipitate and monitor the Fe$^{3+}$ concentration in the solution. As presented in the Fig. S2 in ESI†, the formation of the precipitate has an induction period of 2 hours, at which the Fe$^{3+}$ concentration decreases slightly, followed by an exponential decay, which results in negligible concentration of Fe$^{3+}$ in solution after 9 hours, whilst maintaining a constant pH of 6.6. This experiment led us to find out the minimum time required to precipitate all Fe$^{3+}$ from the starting solution, which is approximately 9 hours.

The precipitate formed after 24 hours of stirring was isolated, washed several times with water, dried at room temperature and analyzed by FT-IR, XRD and TEM (Fig. 5b-d).

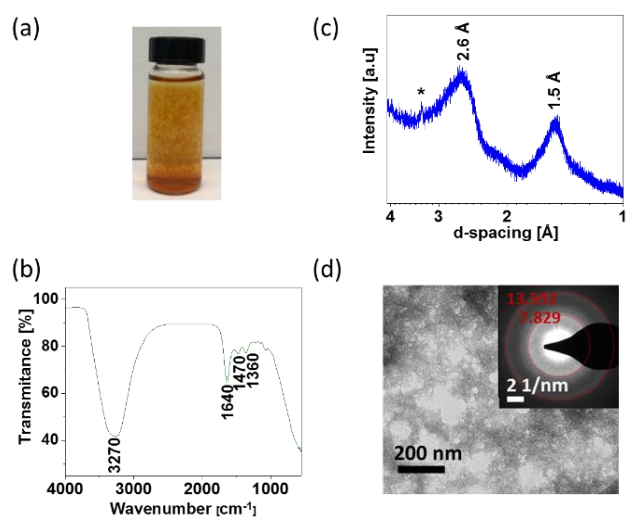

Fig. 4 Reaction mixture after overnight stirring at room temperature (a); FT-IR spectrum (b); XRD pattern (c); TEM micrograph (inset is associated SAED pattern) (d) of the dried precipitate.

In the FT-IR spectrum (Fig. 5b) a broad adsorption band with a maximum at 3270 cm$^{-1}$ is attributed to $\upsilon_{O-H}$ stretching mode, which is typical for hydroxides and also to structural/adsorbed water. The peak at 1640 cm$^{-1}$ additionally confirms the presence of water in the precipitate. Furthermore, the peaks at 1470 cm$^{-1}$ and 1360 cm$^{-1}$, assigned to the Fe-O and Fe-OH vibrations, respectively, suggest that the precipitate is most probably one of the forms of ferric oxyhydroxide.[26, 27] The very strong absorption at wavenumber <1000 cm$^{-1}$ can be explained by the deformation vibrations of surface O-H groups in ferric hydroxides.[27] The weak absorption bands at the region of 1200-1000 cm$^{-1}$ might be related to the partial complexation of Fe$^{3+}$ by TEA molecules.[28]

The XRD pattern of the precipitate (Fig. 5c) shows that the material presents a low degree of crystallinity, as only two broad peaks at about 2.6 Å and 1.5 Å are detected, which are typical for iron oxyhydroxides previously reported.[29-31] Additional characterization by TEM and SAED (Fig. 5d) demonstrate that the material presents an irregular morphology and the particles are strongly agglomerated. The SAED (inset in Fig. 5d) highlights only two bright, diffused rings which positions fit well to the pattern obtained by XRD. TEM results are in a good agreement with those found in the literature suggesting that the precipitate formed at room temperature is most probably one of the forms of ferrihydrite.[32] This thermodynamically metastable $Fe^{3+}$ oxyhydroxides finally transform to more crystalline forms depending on the temperature and pH.[33, 34] Considering that those parameters are both changing during the further thermal decomposition of urea, we hypothesize that, in presence of $Ni^{2+}$ and $CO_3^{2-}$, the ferrihydrite transforms into Ni-Fe LDH.

Since most of the $Fe^{3+}$ precipitates as ferrihydrite, the remaining solution is expected to be composed of $Ni^{2+}$, $NO_3^-$, TEA and urea. In order to understand the coordination of $Ni^{2+}$, the reaction mixture was analyzed by NMR and UV-Vis spectroscopy and compared to the spectra of pure $Ni(NO_3)_2$ and TEA (Fig. 6).

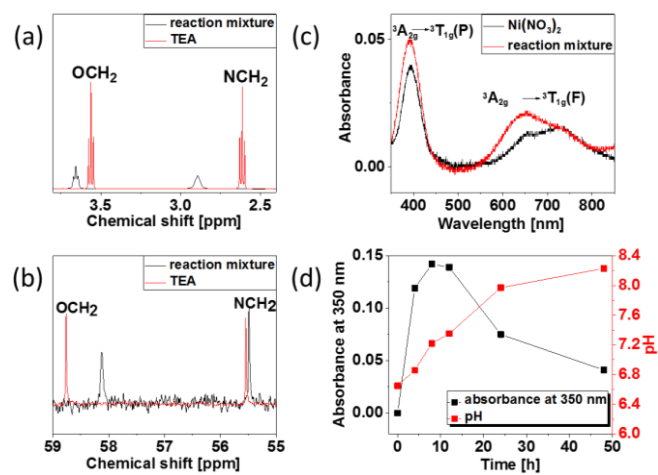

Fig. 5 $^1$H NMR and $^{13}$C NMR spectra of the reaction mixture and TEA (a, b); UV-Vis spectra of the reaction mixture and $Ni(NO_3)_2$ (c); absorbance at 350 nm related to $Fe^{3+}$ concentration and pH value during 48 hours of heating the reaction mixture at 100 °C (d).

Firstly, the reaction mixture was characterized by $^1$H and $^{13}$C NMR spectroscopy and compared to the spectrum of pure TEA (Fig. 6a, b). The changes in both $^1$H and $^{13}$C environment are clearly visible: $^1$H NMR spectrum of pure TEA shows two triplets for the methylene groups bonded to N and O, respectively, which are observed to be broadened and shifted downfield in the reaction mixture due to the presence of $Ni^{2+}$. This is in good agreement with the theoretical predictions, suggesting that the loss of electron-density upon coordination to the metal results in a shift to higher frequency of the protons adjacent to the ligand.[35] In contrast, in the $^{13}$C NMR spectrum the signals are shifted upfield, which was also previously observed in metal-organic compounds,[36] which is an additional proof of the formation $Ni^{2+}$-TEA complex. Several coordination compounds of Ni-TEA-$NO_3^-$ have been described in the form of blue crystals,[37, 38] which differs from the clear, pale green/blue solution obtained in the present study. This difference is most probably the result of an insufficient amount of TEA in relation to $Ni^{2+}$, since the TEA:$Ni^{2+}$ ratio is 4:3 instead of 2:1 to form $[Ni(TEA)_2](NO_3)_2$ in a solid form.

The UV-Vis absorption spectra of both $Ni(NO_3)_2$ and the reaction mixture (Fig. 6c) show two weak bands assigned to d-d transitions corresponding to an octahedral coordination geometry of $Ni^{2+}$.[37, 39, 40] A sharp peak centred at 390 nm can be assigned to $^3A_{2g} \rightarrow {}^3T_{1g}(P)$ transition in both cases,[37, 39] while the absorbance is higher in case of the reaction mixture, which probably is a result of metal coordination by TEA ligands. Moreover, the doublet with maxima at 656 nm and 726 nm ascribed to $^3A_{2g} \rightarrow {}^3T_{1g}(F)$ transition typical of nickel(II) aqua complex ($[Ni(H_2O)_6]^{2+}$)[40, 41] is slightly shifted towards lower wavelengths in the reaction mixture. Higher absorbance of the peak at 390 nm and a blue shift at 656-720 nm were previously observed by Agarwala et al.[41] for various Ni-TEA complexes in solution including $[Ni(TEA)_2]^{2+}$, $[Ni(TEA)_2(OH)_2]^{2+}$ and $[Ni(TEA)]^{2+}$.

Therefore, on the basis of UV-Vis and NMR spectroscopy, nickel(II) is most probably present in the reaction mixture as an octahedral Ni-TEA metal complex, however the chemical formula cannot be accurately defined, due to many possibilities of ligand coordination,[41, 42] and most probably it consists as a mixture of different Ni-TEA derivatives in quick kinetic equilibrium.

*Step 2- Heating to 100 °C*

The heating of the reaction mixture (gel-like ferrihydrite and $Ni^{2+}$-TEA complex) to 100 °C was monitored by UV-Vis measurements (Fig. 6d). Within the first 4 hours of heating the mixture, a dramatic raise of $Fe^{3+}$ concentration was observed, suggesting the decomposition of ferrihydrite. The $Fe^{3+}$ concentration augmented for a maximum of 8 hours and started decreasing until the end of the measurements. Nevertheless, after 48 hours of heating the reaction mixture, some unbounded $Fe^{3+}$ ions remained in the solution, as observed in Figure 6d. This is consistent with the $Ni^{2+}/Fe^{3+}$ ratio in the final product (3.54), which is slightly higher than initially used (3.00) suggesting that not all $Fe^{3+}$ was incorporated to the Ni-Fe LDH.

The evolution of the pH is also plotted in Fig. 6d. During the first 12 hours, a linear increase of the pH is observed due to the urea hydrolysis at 100 °C. After this time, the consumption of $OH^-$ by metal cations leads to a slower pH increase. Carbonate anions produced from urea decomposition intercalate between the layers to neutralize a positive charge generated by $Fe^{3+}$ substitution within $Ni(OH)_2$.

The precipitates formed after 4, 12 and 24 hours of heating were analysed by SEM (Fig. S3 in ESI†), in order to investigate the platelets formation process. After 4 and 12 hours, hexagonal flakes with slightly jagged edges and diameters around 0.5 μm mixed with an amorphous precipitate (most probably, remaining ferrihydrite) were observed. Further heating for an additional 12 hours lead to the complete consumption of the amorphous precipitate and only hexagonal platelets remained. Nevertheless, the heating was continued for additional 24 hours, in order to maximize the incorporation of $Fe^{3+}$ to the LDH.

## Conclusion

In summary, we obtained Ni-Fe LDH hexagonal platelets by a homogeneous precipitation method at 100 °C under atmospheric pressure. The flakes' diameter is in the range of 0.5-1.5 μm and thicknesses between 15 and 20 nm. This material is highly crystalline and the metal cations are homogeneously distributed within a single flake. We believe that those platelets are formed in a two step process, where firstly reaction intermediates (iron oxyhydroxide and Ni-TEA complex mixture) are formed at room temperature, and then, upon heating to 100 °C, $Ni^{2+}$ and $Fe^{3+}$ are released from their precursors and react with $OH^-$ and $CO_3^{2-}$ (from urea hydrolysis) forming regular hexagons of Ni-Fe LDH. The obtained material showed favourable electrocatalytic activity towards OER, exhibiting an overpotential of 0.36 V vs RHE, which is consistent with literature reports.

We have demonstrated in this paper that high quality $Fe^{3+}$-containing LDH platelets can be produced by a simple method at 100 °C in a conventional synthesis setup.

## Experimental

**Chemicals.** Nickel(II) nitrate hexahydrate ($Ni(NO_3)_2 \cdot 6H_2O$, extra pure, Fisher Scientific), Iron(III) nitrate nonahydrate ($Fe(NO_3)_3 \cdot 9H_2O$, purity <98%, Sigma Aldrich), urea ($CH_4N_2O_2$, purity ≥ 99.3 %, Alfa Aesar) and TEA ($C_6H_{15}NO_3$, purity ≥ 99 %, Sigma Aldrich) were used as received without further purification.

**Synthesis of Ni-Fe LDH hexagonal platelets.** $Ni(NO_3)_2 \cdot 6H_2O$, $Fe(NO_3)_3 \cdot 9H_2O$ and urea were dissolved in 80 ml of deionized water to a final concentration of 7.5, 2.5 and 17.5 mM, respectively. Then, 0.8 mmol of TEA were added and the solution was stirred at room temperature for 24 hours, which led to the formation of gel-like brown precipitate. After that, the reaction mixture was transferred to 100 ml round-bottom flask and immersed in an oil bath previously heated to 100 °C under reflux conditions and heated for 48 hours. The flask was naturally cooled down to room temperature and the obtained precipitate was collected by centrifugation (3000 rpm/ 10 min) and washed with water several times.

**Equipment and characterization techniques.** Scanning electron microscopy (SEM) images were acquired using a *Zeiss Utra Plus* (Carl Zeiss AG, Germany) operated at 2-3 kV, while energy dispersive X-ray spectroscopy (EDX) mapping was performed at 15 kV. Transmission electron microscopy (TEM) and selected area electron diffraction (SAED) were conducted using an *FEI Titan* (FEI, Oregon, USA) microscope operated at 80 keV. Atomic force microscopy (AFM) images were taken on an *Asylum Research MFP 3D* microscope working in a tapping mode. Fourier-transform infrared (FT-IR) spectra were recorded using a *Perkin Elmer Spectrum 100* FT-IR spectrometer via attenuated total reflectance (ATR) method. Powder X-ray diffraction (XRD) was measured in a *Bruker Advance Powder X-ray* diffractometer equipped with a Mo-Kα emission source (λ= 0.7107 Å) in the Bragg-Brentano configuration. The metals content was measured with a *Varian 55 Atomic Absorption* spectrometer. Thermogravimetric analysis (TGA) was carried out in a *Perkin Elmer Pyris 1 TGA* (the samples were heated up to 850 °C at a rate of 10 °C min$^{-1}$) under air atmosphere. Ultraviolet-visible (UV-Vis) spectroscopy measurements were conducted using a *Biochrom Libra S22* UV/Vis Spectrophotometer. $^1H$ and $^{13}C$ nuclear magnetic resonance (NMR) analyses were performed using a *Bruker Avance HD 400 NMR* equipped with a BBFO probe. An ultrasonic spray deposition (*USI Prism Ultracoat 300*) was used to deposit the sample dispersion onto nickel foam substrates (0.25 cm$^2$) and prepare electrodes with an average mass load of 0.3 mg·cm$^{-2}$. Electrochemical measurements were conducted in a *BioLogic VMP 300*. A platinum wire and Ag/AgCl were used as counter and reference electrodes in 1 M KOH electrolyte solution, respectively. Both cyclic and linear sweep voltammetry curves were acquired in the range of 0 V to 0.6 V at a scan rate of 5 mV·s$^{-1}$ against Ag/AgCl. All the potentials were calibrated with respect to a reversible hydrogen electrode (RHE) reference as follows:[2] $V_{RHE} = V_{Ag/AgCl} + V_{Ag/AgCl(vsRHE)} + (0.059 \times pH)$. All voltammetry curves are shown with compensated cell resistance (iR).

## Conflicts of interest

There are no conflicts to declare.

## Acknowledgements

The authors would like to thank the following funding support: Science Foundation Ireland (AMBER), European Research Council (3D2DPrint & TC2D), Horizon 2020 NMP (CoPilot) and 7th Framework Programme (ITN MoWSeS).

# Supplementary information

# Low-temperature synthesis of high quality Ni-Fe layered double hydroxides hexagonal platelets

Sonia Jaśkaniec [a,b], Christopher Hobbs [b,c], Andrés Seral-Ascaso [b,c], João Coelho [b,c], Michelle P. Browne [d], Daire Tyndall [a,b], Takayoshi Sasaki [e] and Valeria Nicolosi [a,b,c]

[a] School of Chemistry, Trinity College Dublin, Ireland.
[b] CRANN&AMBER, Trinity College Dublin, Ireland.
[c] School of Physics, Trinity College Dublin, Ireland.
[d] School of Chemistry, Queens University Belfast, Ireland.
[e] National Institute for Materials Science, Tsukuba, Japan

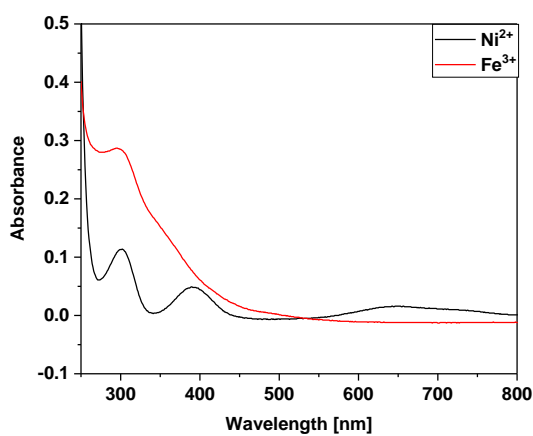

**Fig. S1** UV-Vis spectra of $Ni(NO_3)_2$ and $Fe(NO_3)_3$ dissolved in water.

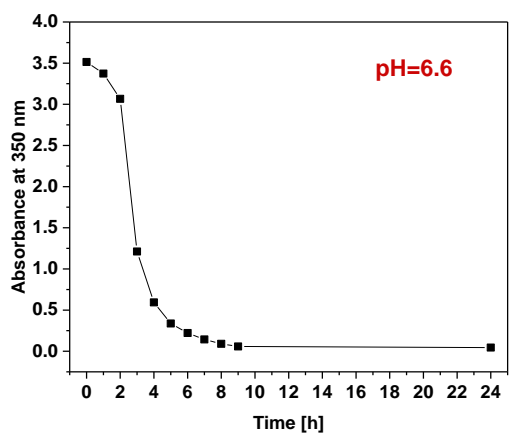

**Fig. S2** Absorbance at 350 nm at different stirring time at room temperature.

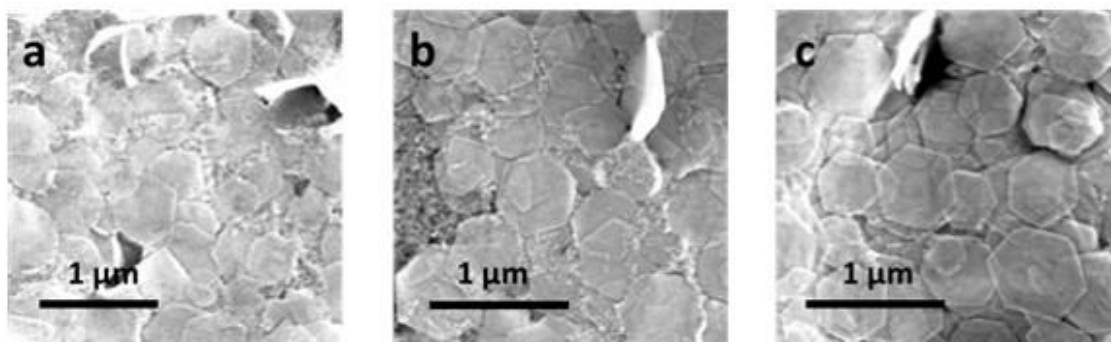

**Fig. S3** SEM micrographs of $Ni_xFe_y$ LDH formed after a) 4 hours, b) 12 hours and c) 24 hours of heating.